\documentclass[aoas,MSNbibl,nameyear,seceqn,dvips]{arximspdf}
\usepackage{dcolumn}
\usepackage{graphicx}

\doi{10.1214/10-AOAS453}
\volume{5}
\issue{2B}
\pubyear{2011}
\firstpage{1407}
\lastpage{1424}

\makeatletter
\newcolumntype{d}[1]{D{.}{.}{#1}}

\def\@bmisc[#1]{%
  \get@battribute{unstr}%
  \common@pub@types%
  \let\bauthor\bbl@bauthor%
  \let\bhowpublished\@firstofone%
  \def\borganization##1{{\bauthor@style ##1}}%
}

\makeatother

\begin{document}
\begin{frontmatter}

\title{Network routing in a dynamic environment\thanksref{TT2,SUPORT}}
\runtitle{Network routing}

\begin{aug}
\author[A]{\fnms{Nozer D.} \snm{Singpurwalla}\corref{}\ead[label=e1]{nozer@gwu.edu}}
\runauthor{N. D. Singpurwalla}
\affiliation{George Washington University}
\address[A]{Department of Statistics\\
George Washington University\\
2140 Pennsylvania Avenue\\
 Washington, District of Columbia 20052\\
  USA\\
  \printead{e1}} 
\end{aug}
\relateddoi{TT2}{Discussed in \doi{10.1214/11-AOAS453A}.}
\thankstext{SUPORT}{Supported by the U.S. Army Research Office under Grant W911NF-09-1-0039 and
by NSF Grant DMS-0915156.}

\received{\smonth{12} \syear{2009}}
\revised{\smonth{12} \syear{2010}}

\begin{abstract}
Recently, there has been an explosion of work on network
routing in hostile environments. Hostile environments tend to be dynamic,
and the motivation for this work stems from the scenario of IED placements
by insurgents in a logistical network. For discussion, we consider here a
sub-network abstracted from a real network, and propose a framework for
route selection. What distinguishes our work from related work is its
decision theoretic foundation, and statistical considerations pertaining to
probability assessments. The latter entails the fusion of data from diverse
sources, modeling the socio-psychological behavior of adversaries, and
likelihood functions that are induced by simulation. This paper demonstrates
the role of statistical inference and data analysis on problems that have
traditionally belonged in the domain of computer science, communications,
transportation science, and operations research.
\end{abstract}


\end{frontmatter}

\section{Introduction: Background and overview}\label{sec1}

Network routing problems involve the selection of a pathway from a source to
a sink in a network. Network routing is encountered in logistics,
communications, the internet, mission planning for unmanned aerial vehicles,
telecommunications, and transportation, wherein the cost effective and safe
movement of goods, personnel, or information is the driving consideration.
In transportation science and operations research, network routing goes
under the label \textit{{vehicle routing problem}} (VRP); see
Bertsimas and Simchi-Levi (\citeyear{Bertsimas1996}) for a survey. The flow of any commodity
within a network is hampered by the failure of one or more pathways that
connect any two nodes. Pathway failures could be due to natural and physical
causes, or due to the capricious actions of an adversary. For example, a
cyber-attack on the internet, or the placement of an improvised explosive
device (IED) on a pathway by an insurgent. Generally, the occurrence of all
types of failures is taken to be probabilistic.
See, for example, Gilbert (\citeyear{Gilbert1959}), or Savla, Temple and Frazzoli (\citeyear{Savla2008}) who assume that the placement
of mines in a region can be described by a spatio-temporal Poisson process.

\begin{figure}
\vspace*{-3pt}
\includegraphics{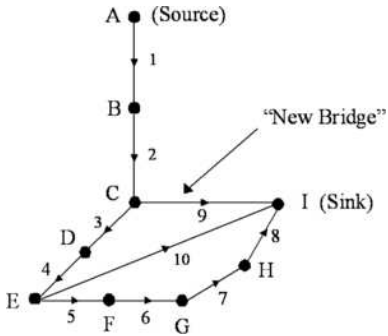}
\vspace*{-3pt}
\caption{Subnetwork for transportation from A to I.}
\label{fig1}
\vspace*{-10pt}
\end{figure}

The traditional approach in network routing assumes that the failure
probabilities are fixed for all time, and known; see, for example,
Colburn (\citeyear{Colburn1987}).\vadjust{\goodbreak} Modern approaches recognize that networks operate in dynamic
environments which cause the failure probabilities to be dynamic. Dynamic
probabilities are the manifestations of new information, updated knowledge,
or new developments (circumstances); de Vries, Roefs and Theunissen (\citeyear{de2007})
articulate this matter for unmanned aerial vehicles.

The work described here is motivated by the placement of IED's on the
pathways of a logistical network; see Figure~\ref{fig1}. Our aim is to prescribe an
optimal course of action that a decision maker $\mathcal{D}$ is to take vis-%
\`{a}-vis choosing a route from the source to the sink. By optimal action we
mean selecting that route which is both cost effective and safe. $\mathcal{D}
$'s efforts are hampered by the actions of an adversary $\mathcal{A}$, who
unknown to $\mathcal{D}$, may place IED's in the pathways of the network. In
military logistics, $\mathcal{A}$ is an insurgent; in cyber security, $%
\mathcal{A}$ is a hacker. $\mathcal{D}$'s uncertainty about IED presence on
a~particular route is encapsulated by $\mathcal{D}$'s personal probability,
and $\mathcal{D}$'s actions determined by a judicious combination of
probabilities and $\mathcal{D}$'s utilities. For an interesting discussion
on a military planner's attitude to risk, see \citeauthor{de2007} (\citeyear{de2007}) who
claim that individuals tend to be risk prone when the information presented
is in terms of losses, and risk averse when it is in terms of gains. Methods
for a meaningful assessment of $\mathcal{D}$'s utilities are not on the
agenda of this paper; our focus is on an assessment of $\mathcal{D}$'s
probabilities, and the unconventional statistical issues that such
assessments spawn.

To cast this paper in the context of recent work in route selection under
dynamic probabilities, we cite Ye et al. (\citeyear{Ye2010}) who consider minefield
detection and clearing. For these authors, dynamic probabilities are a
consequence of improved estimation as detection sensors get close to their
targets. The focus of their work is otherwise different from the decision
theoretic focus of ours.

We suppose that $\mathcal{D}$ is a coherent Bayesian and thus an expected
utility maximizer; see Lindley (\citeyear{Lindley1985}). This point of view has been
questioned by de Vries, Roefs and Theunissen (\citeyear{de2007}) who claim that humans use heuristics to
make decisions. The procedures we endeavor to prescribe are on behalf of $%
\mathcal{D}$. We do not simultaneously model $\mathcal{A}$'s actions, which
is what would be done by game theorists. Rather, our appreciation of $%
\mathcal{A}$'s actions are encapsulated via likelihood functions, and
modeling socio-psychological behavior via subjectively specified
likelihoods is a novel feature of this paper. Fienberg and Thomas (\citeyear{Fienberg2010})
give a nice survey of the diverse aspects of network routing dating from the
1950s, covering the spectrum of probabilistic, statistical, operations
research, and computer science literatures. In Thomas and Fienberg (\citeyear{Thomas2010}) an
approach more comprehensive than that of this paper is proposed; their
approach casts the problem in the framework of social network analysis,
generalized linear models, and expert testimonies.

\vspace*{-4pt}
\subsection{Overview of the paper}\label{sec1.1}

We start Section~\ref{sec2} by presenting a subnetwork, which is part of a real
logistical network in Iraq, and some IED data experienced by this
subnetwork. For security reasons, we are unable to present the entire
network and do not have access to all its IED experience. Section~\ref{sec3}
pertains\vadjust{\goodbreak} to the decision-theoretic aspects of optimal route selection. We
discuss both the nonsequential and the sequential protocols. The latter
raises probabilistic issues, pertaining to the ``Principle of
Conditionalization,''  that appear to have been overlooked
by the network analyses communities. The material of Section~\ref{sec3} constitutes
the general architecture upon which the material of Section~\ref{sec4} rests.
Section~\ref{sec4} is about the inferential and statistical matters that the
architecture of Section~\ref{sec3} raises. It pertains to the dynamic assessment of
failure probabilities, and describes an approach for the integration of data
from multiple sources. Such data help encapsulate the actions of $\mathcal{A%
}$, and $\mathcal{D}$'s efforts to defeat them. The approach of Section~\ref{sec4}
is Bayesian; it entails the use of logistic regression and an unusual way of
constructing the necessary likelihood functions. Section~\ref{sec5} summarizes the
paper, and portrays the manner in which the various pieces of Sections~\ref{sec3} and~\ref{sec4}
 fit together. Section~\ref{sec5} also closes the paper by showing the workings of
our approach on the network of Section~\ref{sec2}.

\vspace*{-4pt}
\section{A network for transportation logistics}\label{sec2}

Figure~\ref{fig1} is a subnetwork abstracted from a real logistics network used in
Iraq. The subnetwork has nine nodes, labeled A (not to be confused with
adversary $\mathcal{A}$) to I, and ten links, labeled 1 to~10. The source
node is A and the sink node is I.

There are thirteen bridges dispersed over the ten links of Figure~\ref{fig1}, with
link 9 having one bridge, the ``\textit{new
bridge}.'' This bridge is a mile away from a park, the old
city, the bus station, and the mosque. The precise locations of the
remaining 12 bridges in the subnetwork are classified. There have been four
crossings on the ``new bridge,''   and none
of these have experienced an IED attack. To plan an optimal route from
source to sink, $\mathcal{D}$ needs to know the probability of experiencing
an IED attack on the next crossing on each of the ten links. However, we
focus discussion on link 9, because it is for this link that we have
information on the number of previous crossings.

To assess the required probabilities, we need to have all possible kinds of
information, including that given in Table~\ref{tab1}, which gives the history of IED
placements on the remaining twelve bridges of the subnetwork. The data of
Table~\ref{tab1}, though public, were painstakingly generated via information from
multiple sources---such as Google Maps---by the so-called process of
``connecting the dots.''   Generally, such
data are hard to come by via the public domain. The recently released
WikiLeaks (\citeyear{WikiLeaks2010}) data has some covariate information on IED experiences in
Afghanistan. However, there are very few well-defined logistical routes in
Afghanistan, and those that may be there are not identified in the WikiLeaks
database. Furthermore, the covariate information that is available is not of
the kind relevant to route selection. Thus, for this paper, the
WikiLeaks--Afghanistan data are of marginal value.

\begin{table}
  \caption{Historical data on IED placements on 12 bridges in Iraq}\label{tab1}
\begin{tabular}{@{}lcd{1.2}d{1.2}d{1.2}d{1.2}@{}}
\hline
\textbf{Bridge} & \multicolumn{1}{c}{\textbf{Attack}} & \multicolumn{1}{c}{\textbf{Park}} & \multicolumn{1}{c}{\textbf{Old city}} & \multicolumn{1}{c}{\textbf{Bus station}} & \multicolumn{1}{c@{}}{\textbf{Mosque}}\\
\hline
Aimma & 0 & 0 & 0 & 1 & 0.1\\
Adhimiya & 0 & 0.25 & 0.75 & 1.5 & 1\\
Sarafiya & 1 & 0 & 1 & 1 & 0.5\\
Sabataash & 0 & 1 & 0 & 0.75 & 0.2\\
Shuhada & 0 & 2 & 0 & 0.75 & 0.1\\
Ahrar & 0 & 1 & 0 & 1 & 0.75\\
Sinak & 0 & 0.5 & 0 & 1 & 0.3\\
Jumhuriya & 0 & 0.1 & 0 & 0.75 & 0.3\\
Arbataash & 1 & 0 & 3 & 3.5 & 2\\
Jadriya & 1 & 0 & 5.5 & 5 & 2\\
SJadriya & 0 & 0 & 6 & 5.5 & 3\\
Dora & 1 & 2 & 5 & 4 & 4\\
\hline
\end{tabular}
\end{table}

In Table~\ref{tab1}, the column labeled ``Attack''
is 1 whenever the bridge has experienced an attack; otherwise it is $0$. The
other columns give the distance of the bridge, in miles, from population
centers like a park, old city, bus station, and mosque. An entry of zero
denotes that the bridge is next to the landmark. Whereas data on IED attacks
tends to be public (because of press reports), data on the number of
crossings by convoys, the number of IEDs cleared, the composition of the
convoys, etc., remains classified.

The three routes suggested by Figure~\ref{fig1} are as follows: $(1,2,3,4,5,6,7,8)$, $%
(1,2,3,4,10)$, and $(1,2,9)$. Since IEDs are placed by adversaries, $%
\mathcal{D}$ is generally uncertain of their presence when planning
begins. Additionally, there are pros and cons with each route in terms of
distance traversed, route conditions (such as the number of curves and
bends, terrain topology), proximity to hostile territory, receptiveness of
the local population to harbor insurgents, and so on. In actuality, $%
\mathcal{D}$ will have access to historical data of the type shown in Table~\ref{tab1}, and also information about the nature of the cargo, the convoy speed,
intelligence about the cunningness and sophistication of the insurgents, the
number of previous unencumbered crossings on a link, etc.\looseness=-1

$\mathcal{D}$'s problem is to select an optimal route between the three
routes given above. A variant is to specify the optimal route
\textit{sequentially}. That is, start by going from A to C via
links 1 and 2, and then, upon arrival at C, make a~decision to proceed along
link 9 to the sink, or to take the circuitous routes via the links 3 to 8,
and 10, to get to the sink. Similarly, upon arrival at node E, $\mathcal{D}$
could proceed along link 10, or via the links $5,6,7,$ and 8 to arrive at the
sink. $\mathcal{D}$'s decision as to which choice to make will be based on $%
\mathcal{D}$'s uncertainty of IED presence on the links 3 to 10, assessed
when $\mathcal{D}$ is at node~C and at node E.

Thus, optimal route selection is a problem of decision under uncertainty.
Because of the dynamic environment in which convoys operate, $\mathcal{D}$'s
uncertainties change over time. In Section~\ref{sec3} we prescribe a
decision-theoretic architecture for route selection. This requires that $%
\mathcal{D}$ assess his (her) uncertainties about IED placements, as well as
utilities for a successful or failed traversal. Since $\mathcal{D}$'s
uncertainties are dynamic, the prescription of Section~\ref{sec3} is also dynamic;
that is, the selected route is optimal only for an upcoming trip. The main
challenge therefore is an assessment of the dynamic probabilities; see
Section~\ref{sec4}.

\section{$\mathcal{D}$'s decision-theoretic architecture}\label{sec3}

\begin{figure}

\includegraphics{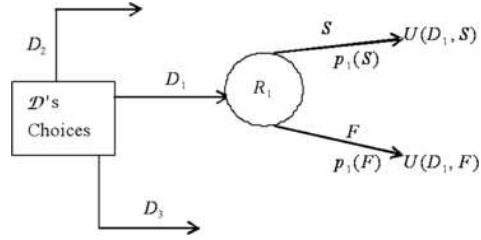}

\caption{$\mathcal{D}$'s decision tree for nonsequential actions.}
\label{fig2}\end{figure}

Under the nonsequential protocol, $\mathcal{D}$ needs to choose, at
decision time, from the following: $D_{1}\equiv $ take route $(1,2,9)$; $D_{2}\equiv $ take
route $(1,2,3,4,5,6,7,8)$; or $D_{3}\equiv $ take route $(1,2,3,4,10)$.
Figure~\ref{fig2} shows $\mathcal{D}$'s decision tree for these choices, with each $%
D_{i}$ leading to a~random node $R_{i}$, with each $R_{i}$ leading to an
outcome $S$ (for success) and $F$ (for failure), $i=1,2$. Here $S$ is the
event that an IED is not encountered on any link of the route, and $F$ the
event that an IED is encountered. If $\mathcal{D}$ is aware of any route
clearing activity, then this becomes a part of $\mathcal{D}$'s covariates
used to assess probabilities. The presence of an IED does not necessarily
imply an explosion. Unexploded IEDs cause disruptions, and $\mathcal{D}$'s
aim is to choose that route which minimizes the risk of damage and
disruption.

In Figure~\ref{fig2}, $p_{1}(S)$ and $p_{1}(F)=1-p_{1}(S)$ denote $\mathcal{D}$'s
probabilities for success and failure, and $U(D_{1},S)$ and $U(D_{1},F)$, $%
\mathcal{D}$'s utilities under $D_{1}$. The quantities $p_{2}(S)$, $p_{2}(F)$%
, $U(D_{2},S)$, and $U(D_{2},F)$ pertain to $D_{2}$; similarly, for $D_{3}$.

Assessing utilities is a substantive task [cf. Singpurwalla (\citeyear{Singpurwalla2010})]
entailing rewards, penalties, and attitudes to risk. This task is not
pursued here. However, one often assumes binary loss functions, so that $%
U(D_{i},S)=1$ and $U(D_{i},F)=0$.

Per the principle of \textit{maximization of expected utility}, $\mathcal{D}$
chooses that $D_{i}$ for which the expected utility is a maximum. Thus, at
each $R_{i}$, $\mathcal{D}$ computes, for $i=1,2,3$,%
\[
\mathbf{E}[U(D_{i})]=p_{i}(S)U(D_{i},S)+p_{i}(F)U(D_{i},F),
\]%
and chooses that $D_{i}$ which maximizes $\mathbf{E}[U(D_{i})]$.

\subsection{$\mathcal{D}$'s assessment of $p_{i}(S)$}\label{sec3.1}

The building blocks of $p_{i}(S)$ are the $p(j)$'s, $\mathcal{D}$'s
probabilities of an IED placement on link $j$, $j=1,\ldots,10$. Under action~$%
D_{1}$, the event $S$ will occur at the terminus of the tree if there is no
IED placement on the links 1, 2, and 9. If $E(j)$ denotes the event that
an IED is placed on link $j$, then $p(j)$ is an abbreviation for $P(E(j))$.
If $\mathcal{D}$ assumes that the $E(j)$'s, $j=1,2,9$, are
independent, then%
\[
p_{1}(S)=\bigl(1-p(1)\bigr)\bigl(1-p(2)\bigr)\bigl(1-p(9)\bigr) \quad    \mbox{and} \quad    p_{1}(F)=1-p_{1}(S);
\]%
otherwise,%
\[
p_{1}(F)=p(1)+p(2)+p(9)-p(1,2)-p(1,9)-p(2,9)+p(1,2,9),
\]%
where $p(j,k)$ is $\mathcal{D}$'s joint probability that both $E(j)$ and $%
E(k)$ occur, $j\neq k$; similarly with $p(j,k,l)$, $j\neq k\neq l$. If $%
p(j|k)$ denotes $\mathcal{D}$'s conditional probability of $E(j)$ given $%
E(k) $, and if $\mathcal{D}$ judges $E(j)$ independent of $E(l)$, given $%
E(k)$, then%
\[
p(j,k,l)=p(j|k)p(k|l)p(l).
\]%
Conditional independence in networks is often invoked when dependence
between $E(j)$ and $E(k)$ matters only when links $j$ and $k$ are
neighbors. Since links 1 and 9 are not neighbors, $\mathcal{D}$ may judge $%
E(1)$ and $E(9)$ independent given $E(2)$.

$\mathcal{D}$'s main task is to assess the probabilities of the type $p(j)$
and $p(j|k)$. The material of Section~\ref{sec4} pertains to this exercise.

\subsection{Decision making under a sequential protocol}\label{sec3.2}

Here, $\mathcal{D}$ starts with a~single choice, namely, getting to node C
via links 1 and 2, and then, upon arriving at C, making one of two choices:
get to the sink via link 9, or via the links 3 through 8, and 10. With three
choices, the decision tree for the sequential protocol will be analogous to
that of Figure~\ref{fig2}, save for the fact that the decision nodes will be at nodes
C and E, instead of being at node A. The rest of the analysis parallels that
described in the material following Figure~\ref{fig2} [cf. Singpurwalla (\citeyear{Singpurwalla2009})], save
for one matter, namely, the caveat of conditionalization.

\subsubsection{The caveat of conditionalization}\label{sec3.2.1}

The \textit{principle of conditionalization} (POC) pertains to probability
assessments of two (or more) events, and the disposition of one of them
becomes known [cf. Singpurwalla (\citeyear{Singpurwalla2006}), page 21, and (\citeyear{Singpurwalla2007})]. It arises because
conditional probabilities are in the \textit{subjunctive mood}. When the
disposition of the conditioning event becomes known, and the POC is upheld,
the probability of the unconditioned event is its previously assessed
conditional probability. When the POC is not upheld, one assesses the
probability of the unconditioned event via a likelihood and Bayes' Law,
using the revealed value of the conditioned event as data. When sequential
routing is done for strategic reasons, socio-psychological issues come into
play, and then it is realistic to assess the probability of the
unconditioned event via a likelihood.

To illustrate the above, consider the scenario of $\mathcal{D}$ choosing a
sequential protocol, and having arrived at node C needs to assess the
quantities $p_{2,9}(S)$ and $p_{2,3}(S)$, where $p_{2,9}(S)$ is the
probability of successfully arriving at the sink via links 2 and 9. If the
POC is upheld, then $p_{2,9}(S)$ is obtained as $P(E^{c}(9)|E^{c}(2))$; $%
E^{c}(2)$ is the probability of no IED presence on link 2. If the POC is not
upheld, then%
\[
p_{2,9}(S)=P ( E^{c}(9);E^{c}(2) ) \propto \mathcal{L} (
E^{c}(9);E^{c}(2) )  \bigl( 1-p(9) \bigr) ,
\]%
where the middle term is $\mathcal{D}$'s likelihood of an IED absence on
link 9, under the sure knowledge of an IED absence on link 2. Similarly with
$p_{2,3}(S)$.

The likelihood is specified by $\mathcal{D}$ and is the price to be paid for
rejecting the POC. Such likelihoods may encapsulate the socio-psychological
considerations that $\mathcal{D}$ chooses to exercise. Since the likelihood
is a weight that $\mathcal{D}$ assigns to a prior probability, $\mathcal{D}$
may upgrade (downgrade) the prior via the likelihood depending on whether
the absence of an IED on link 2 would make the presence of an IED on link 9
more (or less) likely. Here much depends on what $\mathcal{D}$ thinks of the
abilities and resources of insurgents.

\section{Dynamic assessment of link probabilities}\label{sec4}

By link probabilities, we mean unconditional probabilities of the type $%
p(j),  j=1,\ldots ,10$. By a dynamic assessment, we mean an updating of
each $p(j)$ due to additional information that can come in the form of hard
data, expert testimonies, socio-psychological considerations, or new
covariate information. The updating of a $p(j)$ can come into play at any
time, most often at the commencement of each route scheduling session, or in
the case of sequential routing, at any time during the cycle at an
intermediate node. In what follows, we focus on link $j$, and discuss the
assessment of $p(j)$. A dynamic assessment of the conditional probabilities
$p(j|k)$ is discussed in Section~\ref{sec4.4}.

Factors that influence any $p(j)$ would be covariates such as route
topography (the number of bends, curves, bridges, and surface conditions),
convoy size and composition (materials or humans), convoy speed, time of
transport (day or night), weather conditions, political climate, etc. A
second factor would be historical data on IED placements on link $j$, and on
all the other links in the region. Finally, also relevant would be $\mathcal{%
D}$'s subjective view about $p(j)$, encapsulated via a prior.

\subsection{Notation and terminology}\label{sec4.1}

Let $(X=1)$ denote the event that one or more IEDs are placed on link $j$; $%
(X=1)$ is a proxy for $E(j)$, and \mbox{$P(X=1)$} a proxy for $p(j)$. To avoid
cumbersome notation, we will not endow $X$ with the index $j$. Let $%
Z_{1},\ldots,Z_{k}$ be $k$ covariates that influence $p(j)$, and denote these
by the vector $\mathbf{Z}=(Z_{1},\ldots,Z_{k})$; $\mathbf{Z}$ is assumed known
to $\mathcal{D}$. Suppose that there have been $n$ crossings on link $j$,
with $X_{m}=1(0)$ if the $m$th crossing experienced (did not experience) an
IED, $m=1,\ldots,n$. Let $\mathbf{X}=(X_{1},\ldots,X_{n})$ denote the historical
IED experience on link $j$. Assume that $X_{1}=X_{2}=\cdots =X_{n}=0$, or
that $X_{1}=X_{2}=\cdots =X_{n-1}=0$, and that $X_{n}=1$. That is, $\mathcal{%
D}$ has observed a series of $n$ successes on link $j$, or has just
experienced a failure. Motivation for these extreme cases is given later.

The IED experience for the entire region is in matrix $\mathbf{D}$, where%
\[
\mathbf{D}= \left[
\matrix{
Y_{1} & Z_{11} & \cdots & Z_{k1} \cr
Y_{2} & Z_{12} & \cdots & Z_{k2} \cr
\vdots &  &  &  \cr
Y_{l} & Z_{1l} & \cdots & Z_{kl} \cr
\vdots &  &  &  \cr
Y_{s} & Z_{1s} & \cdots & Z_{ks}
}
 \right] .
\]%
In the $l$th row of $\mathbf{D}$, $Y_{l}=1(0)$ if an IED presence has been
encountered (not encountered) under condition $Z_{1l},\ldots,Z_{kl}$, for $%
l=1,\ldots,s$. Thus, at disposal to $\mathcal{D}$ are the $s$ IED related
experiences in the region, and associated with each experience are the
values of the $k$ covariates that influence each experience. To avoid any
duplicate weighting of data, $\mathbf{X}$ will \textit{not} be a part of $%
\mathbf{D}$. The motivation for excluding $\mathbf{X}$ from $\mathbf{D}$ is
to give link $j$ a special emphasis by incorporating the effect of $\mathbf{X%
}$, which is specific to link $j$, in a vein that is different from $\mathbf{%
D}$.

Let $x_{i}$ be the realization of $X_{i}$, and $y_{l}$ of $Y_{l}$, $i=1,\ldots,n
$ and $l=1,\ldots,s$. Each $x_{i}=1$ or $0$; similarly, $y_{l}$. $\mathbf{D}$
is assumed known to $\mathcal{D}$; its elements may not be controlled by $%
\mathcal{D}$.

$\mathcal{D}$'s task is to assess $P ( X=1;\mathbf{x},\mathbf{Z,D}^{\ast
} ) $, where $\mathbf{x}=(x_{1},\ldots,x_{n})$, and $\mathbf{D}^{\ast }$
is $\mathbf{D}$ with the $Y_{l}$'s replaced by $y_{l}$, $l=1,\ldots,s$. The
above expression is $\mathcal{D}$'s probability of an IED presence on link $%
j $, knowing $\mathbf{x}$, $\mathbf{Z}$, and $\mathbf{D}^{\ast }$.
Assessing this probability is tantamount to fusing data from two sources: IED
experience on link $j$, and historical IED experience in the region wherein $%
j$ resides. It is a form of weighting wherein one borrows strength based on
individual and population characteristics.

\subsubsection{The proposed model}\label{sec4.1.1}

Start by assuming $\mathbf{x}$ unknown, so that $P ( X=1;\mathbf{x},%
\mathbf{Z,D}^{\ast } ) $ is $P(X=1|\mathbf{X;Z,D}^{\ast })$, and invoke
the law of total probability to write
\[
P(X=1|\mathbf{X;Z,D}^{\ast })=\int ^{1}_{0}P(X=1|p,%
\mathbf{X;Z,D}^{\ast })\pi (p|\mathbf{X;Z,D}^{\ast })\,dp,
\]%
where $p$ is a \textit{propensity} [see Singpurwalla (\citeyear{Singpurwalla2006}), page 50], and $%
\pi (p|\mathbf{X;Z,D}^{\ast })$ is $\mathcal{D}$'s uncertainty about $p$,
given $\mathbf{X}$, with $\mathbf{Z}$ and $\mathbf{D}^{\ast }$ known. The
propensity of event $\mathcal{E}$ is the proportion of times $\mathcal{E}$
occurs in an infinite number of trials.

If we assume that, given $p$, the event $ ( X=1 ) $ is independent
of $\mathbf{X}, \mathbf{Z}$, and $\mathbf{D}^{\ast }$,
then
\begin{equation}
P(X=1|\mathbf{X;Z,D}^{\ast })=\int ^{1}_{0}p\cdot \pi (p|%
\mathbf{X;Z,D}^{\ast })\,dp,  \label{eq4.1}
\end{equation}%
and by Bayes' Law,%
\begin{eqnarray*}
\pi (p|\mathbf{X;Z,D}^{\ast }) &\propto &\pi (\mathbf{X}|p\mathbf{;Z,D}%
^{\ast })\cdot \pi (p;\mathbf{Z,D}^{\ast }) \\
&=&\pi (\mathbf{X}|p)\cdot \pi (p;\mathbf{Z,D}^{\ast }),
\end{eqnarray*}%
if given $p$, $\mathbf{X}$ is independent of $\mathbf{Z}$ and $\mathbf{D}%
^{\ast }$. Here $\pi (p;\mathbf{Z,D}^{\ast })$ is $\mathcal{D}$'s
uncertainty about $p$ in light of $\mathbf{Z}$ and $\mathbf{D}^{\ast }$,
and $\pi (\mathbf{X}|p)$ is $\mathcal{D}$'s probability model for $\mathbf{X}
$. Equation (\ref{eq4.1}) now becomes%
\begin{equation}
P(X=1|\mathbf{X;Z,D}^{\ast })\propto \int ^{1}_{0}p\cdot
\pi (\mathbf{X}|p)\cdot \pi (p\mathbf{;Z,D}^{\ast })\,dp.  \label{eq4.2}
\end{equation}%
However, $\mathbf{X}$ is observed as $\mathbf{x}$, and, thus, a probability
model for $\mathbf{X}$ does not make sense. We therefore write $P(X=1|%
\mathbf{X;Z,D}^{\ast })$ as $P(X=1;\mathbf{x,Z,\mathbf{D}^{\ast })}$, and $%
\pi (\mathbf{X}|p)$ as $\mathcal{L}(p;\mathbf{x})$, the likelihood of $p$
under $\mathbf{x}$. Now equation (\ref{eq4.2}) becomes%
\begin{equation}
P(X=1;\mathbf{x,Z,D}^{\ast })\propto \int ^{1}_{0}p\cdot
\mathcal{L}(p;\mathbf{x})\cdot \pi (p;\mathbf{Z,D}^{\ast })\,dp.  \label{eq4.3}
\end{equation}%
Equation (\ref{eq4.3}) is our proposed model for assessing $p(j)$. To proceed, $%
\mathcal{D}$ needs to specify the likelihood $\mathcal{L}(p;\mathbf{x})$ and
$\pi (p;\mathbf{Z,D}^{\ast })$, the posterior of $p$.

\subsection{Psychological considerations in specifying likelihoods}\label{sec4.2}

The IED scenario entails special considerations for specifying $\mathcal{L}%
(p;\mathbf{x})$. These arise because $\mathcal{D}$ needs to incorporate an
insurgent's \textit{socio-psychological behavior} in the IED
placement process, and also $\mathcal{D}$'s strategy for outfoxing the
insurgent.

Recall that with $\mathbf{x}=(0,\ldots,0)$ or $\mathbf{x}=(0,\ldots,0,1)$, the
conventional likelihood of $p$ would be $\mathcal{L}(p;\mathbf{x})=p^{\sum
x_{i}}(1-p)^{n-\sum x_{i}}$, which for the aforementioned $\mathbf{x}$ would
be $(1-p)^{n}$ or $(1-p)^{n-1}\cdot p$. The motivation for the conventional
specification is that a preponderance of failures (i.e., non-IED placements)
should decrease the propensity of an IED placement, and vice versa.
However, the conventional approach, though appropriate for scenarios which
are nonadversarial, is inappropriate for IED placement which embodies an
adversary with a \textit{socio-psychological} \textit{agenda}. It seems that here a preponderance of failures should eventually
increase the propensity of success. Insurgents are opportunistic adversaries
who may allow a series of successful link crossings only to impart to $%
\mathcal{D}$ a sense of false security, while all the time preparing to do
damage on the next crossing. Similarly, an astute $\mathcal{D}$ would view
the occurrence of a success that is preceded by a sequence of failures (i.e.,
non-IED placements) with much pessimism, as a dramatic change in the
operating environment. Essentially, $\mathcal{D}$ would downgrade the
impact of the observed sequence of $(n-1)$ failures and strongly weigh the
impact of the last success. With the above behavioristic considerations,
our proposed likelihood for~$p$, for $\mathbf{x}=(x_{1},\ldots,x_{n})$ fixed,
is of the form%
\[
\mathcal{L}(p;\mathbf{x})=(1-p)^{\sqrt[n]{n-\sum x_{i}}}\cdot p^{\sum x_{i}}.
\]%
When $\mathbf{x}=(0,\ldots,0)$, the above likelihood becomes
\begin{equation}
\mathcal{L}(p;\mathbf{x})=(1-p)^{\sqrt[n]{n}},  \label{eq4.4}
\end{equation}%
and when $\mathbf{x}=(0,\ldots,0,1)$, it is%
\begin{equation}
\mathcal{L}(p;\mathbf{x})=(1-p)^{\sqrt[n]{n-1}}\cdot p .  \label{eq4.5}
\end{equation}

\begin{figure}

\includegraphics{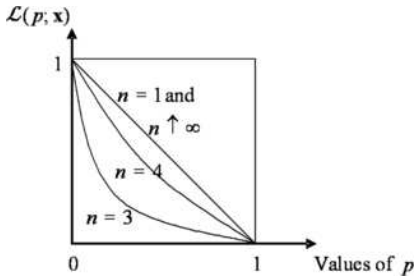}

\caption{The likelihood of $p$ as a function of $n$.}
\label{fig3}\end{figure}

As $n\rightarrow \infty $, equation (\ref{eq4.4}) tends to $(1-p)$, the conventional
likelihood for a single Bernoulli trial that results in a failure. With $%
n\rightarrow \infty $, equation (\ref{eq4.5}) tends to $(1-p)\cdot p$, the
conventional likelihood for the case of two Bernoulli trials resulting in
one failure and one success. In an adversarial context, this is tantamount to
$\mathcal{D}$ regarding a long series of failures as only a single failure
(i.e., $\mathcal{D}$ does not become complacent), and a long series of
failures followed by a success as only one failure and one success. In this
latter case, $\mathcal{D}$ gives equal weight to the $(n-1)$ failures and
the one success; that is, $\mathcal{D}$ becomes deeply concerned when the first
success is observed. Figure~\ref{fig3} illustrates the likelihood.

The proposed likelihood of $p$ is in the envelope bounded by $(1-p)$ and $%
(1-p)^{\sqrt[3]{3}}$. Thus, after three successive failures $\mathcal{D}$
gives more and more weight to larger values of $p$, suggesting an absence of
$\mathcal{D}$'s complacence with a long series of failures. The
specification of the likelihoods as embodied in equations (\ref{eq4.4}) and (\ref{eq4.5}) is
a novel feature of this paper; it is a possible approach to  \textit{adversarial modeling}.

\subsection{$\mathcal{D}$'s assessment of the posterior $\protect\pi (p;%
\mathbf{Z,D}^{\ast })$}\label{sec4.3}

An assessment of the posterior of $p$ in the light of known covariates $%
\mathbf{Z}$ and the historical data $\mathbf{D}^{\ast }$ is developed in two
stages. The challenge here is with the specification of the
likelihood.

\textit{Stage I: Logistic regression for extracting the
information in $\mathbf{D}^{\ast}$.}
Information provided by $\mathbf{D}^{\ast }$ lies in an assessment of the
posterior of $\bolds{\beta }=(\beta _{1},\ldots,\beta _{l},\ldots,\beta _{k})$,
where $\bolds{\beta }$ appears in a logistic regression model%
\[
P(Y_{l}=1;\bolds{\beta },\mathbf{Z}_{l})=\frac{1}{1+\exp  ( -\sum ^{k}_{u=1}Z_{lu}\beta _{u} ) }
\]%
for $l=1,\ldots,s$, with $\mathbf{Z}_{l}=(Z_{1l},\ldots,Z_{kl})$. Recall, $Y_{l}$
and $\mathbf{Z}_{l}$ are the $l$th row of~$\mathbf{D}^{\ast }$.

Using standard but computationally intensive simulation procedures, we can
obtain the posterior of $\bolds{\beta }$ in light of $\mathbf{D}^{\ast
} $. Denote this posterior as $\pi (\bolds{\beta} ;\mathbf{D}^{\ast })$.

\textit{Stage II: The likelihood of ${\large p}$ under $\mathbf{Z}$ and $\mathbf{D}^{\ast }$.}
To assess the posterior $\pi (p;\mathbf{Z,D}^{\ast })$, invoke Bayes' Law to
write%
\begin{equation}
\pi (p;\mathbf{Z,D}^{\ast })\propto \mathcal{L}(p;\mathbf{Z,D}^{\ast })\pi
(p),  \label{eq4.6}
\end{equation}%
where $\mathcal{L}(p;\mathbf{Z,D}^{\ast })$ is the likelihood of $p$ in
light of the known $\mathbf{Z}$ and $\mathbf{D}^{\ast }$, and $\pi (p)$ is $%
\mathcal{D}$'s prior for $p$. Note that $p$ and $\mathbf{Z}$ are specific to
link $i$, whereas $\mathbf{D}^{\ast }$ is common to all the links of the
network. The prior on $p$ could be any suitable distribution, such as a
beta distribution over $(0,1)$. The main theme of Stage II, however, is a
development of the likelihood $\mathcal{L}(p;\mathbf{Z,D}^{\ast })$.

Whereas likelihoods may be subjectively specified, the conventional me\-thod
is to invert a probability model by juxtaposing the parameter(s) and the
random variables. This is the strategy we use, but to do so we need a~%
probability model for $p$ with $\mathbf{Z}$ and $\mathbf{D}^{\ast }$ as
background information. Since~$p$ depends on~$\mathbf{Z}$, we denote this
dependence by replacing $p$ with $p(\mathbf{Z})$. Thus, we seek a~probability
model for $p(\mathbf{Z})$ with $\mathbf{D}^{\ast }$ as a background, namely, $%
P[p(\mathbf{Z});\mathbf{D}^{\ast }]$. But knowing $\mathbf{D}^{\ast }$ is
equivalent to knowing $\bolds{\beta }$ with its posterior probability, $\pi
(\bolds{\beta} ;\mathbf{D}^{\ast })$, developed in Stage I. Thus, for $\bolds{%
\beta }=\bolds{\beta }^{\ast }$, $[p(\mathbf{Z});\bolds{\beta }^{\ast }]$ has
probability $\pi (\bolds{\beta }^{\ast };\mathbf{D}^{\ast })$. However, per
the logistic regression model,%
\[
\lbrack p(\mathbf{Z});\bolds{\beta }^{\ast }]=\frac{1}{1+\exp  ( -%
\sum _{u}^{k}Z_{u}\beta _{u}^{\ast } ) },
\]%
where $\beta _{u}^{\ast }$ appears as the $u$th element of $\bolds{\beta }%
^{\ast }=(\beta _{1}^{\ast },\ldots,\beta _{k}^{\ast })$.

To summarize, the event $[p(\mathbf{Z});\bolds{\beta }^{\ast }]=1/[1+\exp
 ( -\sum Z_{u}\beta _{u}^{\ast } ) ]$ has probability $\pi (\bolds{%
\beta }^{\ast };\mathbf{D}^{\ast })$, and this provides us with a
probability model for $[p(\mathbf{Z});\mathbf{D}^{\ast }]$. Consequently, a
plot of $(p(\mathbf{Z});\bolds{\beta }^{\ast })$ versus $\pi (\bolds{\beta
}^{\ast };\mathbf{D}^{\ast })$ provides the required likelihood function.

\begin{figure}[b]

\includegraphics{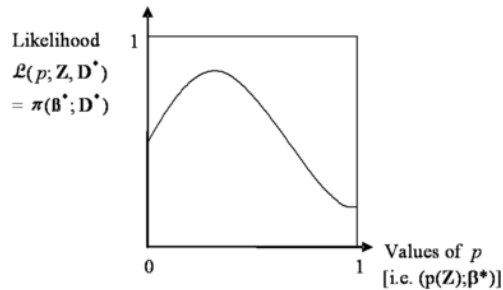}

\caption{The likelihood of $p$ with $\mathbf{Z}$ and $\mathbf{D}^{\ast }$
known.}
\label{fig4}\end{figure}

To implement this idea, we sample a $\bolds{\beta }^{\ast }$ from $\pi (%
\bolds{\beta} ;\mathbf{D}^{\ast })$ to obtain%
\[
\lbrack p(\mathbf{Z});\bolds{\beta }^{\ast }]=\frac{1}{1+\exp  ( -%
\sum _{u}^{k}Z_{u}\beta _{u}^{\ast } ) },
\]%
and also $\pi (\bolds{\beta }^{\ast };\mathbf{D})$. A plot of $(p(\mathbf{Z%
});\bolds{\beta }^{\ast })$ versus $\pi (\bolds{\beta }^{\ast };\mathbf{D}%
^{\ast })$ is then the likelihood function of $p$ in light of $\mathbf{Z}
$ and $\mathbf{D}^{\ast }$; see Figure~\ref{fig4}.

With $\pi (p(\mathbf{Z}))$ the prior on $p$ specified, and the likelihood $%
\mathcal{L}(p;\mathbf{Z,D}^{\ast })$ induced via a logistic regression model
governing $p(\mathbf{Z})$ and $\mathbf{D}^{\ast }$, the desired posterior%
\[
\pi (p;\mathbf{Z,D}^{\ast })\propto \mathcal{L}(p;\mathbf{Z,D}^{\ast })\cdot
\pi (p(\mathbf{Z}))
\]%
can be numerically assessed.

Once the above is done, all the necessary ingredients for obtaining equation
(\ref{eq4.3}), which can now be written as%
\begin{equation}
P(X=1;\mathbf{x,Z,D}^{\ast })\propto \int p\cdot \mathcal{L}(p;\mathbf{x}%
)\cdot \mathcal{L}(p;\mathbf{Z,D}^{\ast })\cdot \pi (p)\,dp,  \label{eq4.7}
\end{equation}%
are at hand. The above expression can be numerically evaluated.

\subsection{Dynamic assessment of conditional probabilities}\label{sec4.4}

For both the nonsequential and sequential protocols wherein the POC is
upheld, we need to assess conditional probabilities of the type $P(n|m)$,
where links $m$ and $n$ are adjacent to each other, and traversing on $m$
precedes that on $n$. There are two possible strategies. The first one is
for $\mathcal{D}$ to subjectively change the assessed $p(n)$ by either
increasing it because an insurgent might find it easy to populate
neighboring links with IEDs, or to decrease it if $\mathcal{D}$ thinks that
an insurgent has limited resources for placing IEDs.

The second approach is less subjective because it incorporates data on IED
placements or nonplacements on neighboring links. The idea here is to treat
the conditioning event $E(m)$ as a covariate, so that the vectors $\mathbf{Z}
$ and $\bolds{\beta }$ of Sections~\ref{sec4.1} and~\ref{sec4.3} get expanded by an
additional term, as $\mathbf{Z}= ( Z_{1},\ldots ,Z_{k},1 ) $ and $%
\bolds{\beta }= ( \beta _{1},\ldots ,\beta _{k},\beta _{k+1} ) $.
Correspondingly, the matrix $\mathbf{D}$ of Section~\ref{sec4.1} also gets expanded
to include an additional column whose $l${{th}} term $Z_{(k+1)l}$ is $%
1 $ whenever there has been an IED experience in a preceding link; otherwise
$Z_{(k+1)l}$ is $0$. With the above in place, a repeat of the exercise
described in Section~\ref{sec4.3} would enable a formal assessment of the conditional
probabilities. The only other matter that remains to be addressed pertains
to the likelihood of $p$ as discussed in Section~\ref{sec4.2}. Since the likelihood
is a~weight assigned to the posterior of $p$, $\mathcal{D}$ may either
increase the $\mathcal{L}(p;\mathbf{x})$ of equations (\ref{eq4.4}) and (\ref{eq4.5}), or
decrease it depending on what $\mathcal{D}$ thinks of an insurgent's
abilities and resources. $\mathcal{D}$ would increase $\mathcal{L}(p;\mathbf{%
x})$ if $\mathcal{D}$ feels that the insurgent's resources are plentiful;
otherwise $\mathcal{D}$ downgrades $\mathcal{L}(p;\mathbf{x})$.

\section{Summary and conclusions}\label{sec5}

Equation (\ref{eq4.7}) shows how $\mathcal{D}$ can assess $p(j)$, the probability of
one or more IED placements on link $j$ in a unified manner by a systematic
application of the Bayesian approach. It entails a fusion of information on
past IED experience on link $j$ (encapsulated by $\mathbf{X}$), historical
data on IED experience in the region (encapsulated by the matrix $\mathbf{D}%
^{\ast }$), and $\mathcal{D}$'s subjective views about $p(j)$, encapsulated
via the likelihood $\mathcal{L}(p;\mathbf{x})$ and the prior $\pi (p)$. The
essence of equation (\ref{eq4.7}) is that its right-hand side is the expected value
of a weighted prior distribution of $p$. The weighting of the prior is by
the product of two likelihoods, one reflecting historical IED experience
specific to link $i$, and the other reflecting historical IED experience in
the region as well as the relevant covariates specific to the forthcoming
trip contemplated by $\mathcal{D}$. The entire development being grounded in
the calculus of probability is therefore \textit{coherent}.

Though cumbersome to plough through, there are novel features to the two
likelihoods. The first likelihood---equations (\ref{eq4.4}) and (\ref{eq4.5})---is an
unconventional likelihood for use with Bernoulli trials. It is motivated by
socio-psychological considerations attributed to both the insurgents who
place the IED's, as well as to $\mathcal{D}$, who does not become complacent
upon a sequence of successful crossings and who upon the occurrence of the
first failure adopts the posture of extreme caution. The second likelihood---that of Figure~\ref{fig4}---is induced in an unusual manner by leaning on the
posterior distribution of the parameter vector of a logistic regression.

The approach of Section~\ref{sec4} displays the manner in which information from
different sources can be fused by decomposing the likelihood of $p$.
Equation (\ref{eq4.7}) shows this. The material of Section~\ref{sec4} feeds into that of
Section~\ref{sec3} which pertains to sequential and nonsequential decision making
under uncertainty.

The computational and simulation work spawned by Section~\ref{sec4} entails logistic
regression, generating $k$-dimensional samples from the posterior
distribution of $\bolds{\beta }$, numerically assessing $\pi (p;\mathbf{Z,D}%
^{\ast })$---equation (\ref{eq4.6})---and numerical integration to obtain $P(X=1;%
\mathbf{x,Z,D}^{\ast })$---equation (\ref{eq4.7}). None of these pose any
obstacles. Section~\ref{sec4.4} pertains to conditional probabilities. It expands on
Sections~\ref{sec4.1} through~\ref{sec4.3}, by treating the conditioning events as
covariates.

\vspace*{-6pt}
\subsection{Data and information requirements}\label{sec5.1}

The one major obstacle pertains to the paucity of the data for validating
the approach. The required data, namely, $\mathbf{x}$, $\mathbf{Z}$, and $%
\mathbf{D}^{\ast }$, are available to the military logisticians, but are
almost always classified. The WikiLeaks data tend to focus on IED explosions
and not on success stories wherein IED's get cleared, similarly with other
publicly available data. Information that is relevant to constructing the
likelihood based on socio-psychological considerations is highly
individualized, and perhaps not even recorded. It is \textit{desirable} to
collect this kind of information via experiments pertaining to the
psychology of logisticians and route planners, and also insurgents via what
is known as ``red teaming.''

The text of this paper can be seen as a template for addressing network
routing in a dynamic environment. The network architecture of Figure~\ref{fig1}
brings out the necessary caveats that problems of this type pose, one such
caveat being the caveat of conditionalization, discussed in Section~\ref{sec3.2.1}.
Real logistical networks are more elaborate. In actual practice the matrix~$%
\mathbf{D}^{\ast }$ could have a very large dimension and thus be
unmanageable. However, given the role that $\mathbf{D}^{\ast }$ plays, one
may simply sample from a high dimensional $\mathbf{D}^{\ast }$ to work with
a more manageable matrix. Besides a prior for $p$, $\pi (p)$, all that is
required of $\mathcal{D}$ are the utilities mentioned in Section~\ref{sec3}.
However, these utilities are proxies for costs, and no form of optimization
can be achieved without cost considerations. Finally, this paper shows how
statistical methodologies can be constructively brought to bear in network
routing problems which generically belong in the domain of computer science,
network analysis, and operations research.

We close this paper by illustrating in Section~\ref{sec5.1} the workings of Sections~\ref{sec3} and~\ref{sec4}
 by using the data of Table~\ref{tab1} to assess the probability of
encountering an IED on the next crossing on the ``new
bridge.''

\subsection{The logistics network revisited}\label{sec5.2}

With respect to the network of Figure~\ref{fig1}, the data of Table~\ref{tab1} maps to the
matrix $\mathbf{D}^{\ast }$ of Section~\ref{sec4.1}, with its column 2 corresponding
to $Y_{l}, l=1,\ldots ,12,$ column 3 corresponding to $Z_{1,1},\ldots
,Z_{1,12},$ and so on, with column 6 corresponding to $Z_{4,1},\ldots
,Z_{4,12}$.

A logistic regression model%
\[
P(Y_{l}=1;\bolds{\beta },\mathbf{Z}_{l})=\frac{1}{1+\exp  ( -\sum ^{4}_{u=0}Z_{ul}\beta _{u} ) }
\]%
for $l=1,\ldots ,12$, with $Z_{0l}=1$, was fitted to the data of Table~\ref{tab1}
using independent Gaussian priors with means $0$ and standard deviations $10$.
This choice of priors is arbitrary. The joint posterior distribution of $%
(\beta _{0},\ldots ,\beta _{4})$ was obtained via Gibbs sampling with 10,000
simulations after a burn-in of 1,000 simulations.

The marginal posterior distributions of $\beta _{0},\beta _{2}$, and $\beta
_{4}$ were symmetric looking, but those of $\beta _{1}$ and $\beta _{3}$
were skewed to the left; plots of these distributions are not shown. Table~\ref{tab2}
compares posterior means against their maximum likelihood estimates, showing
a good agreement between the two, save for $\beta _{0}$.

\begin{table}[b]
  \caption{Comparison of Bayes' versus maximum likelihood estimates}\label{tab2}
\begin{tabular}{@{}lccccc@{}}
   \hline
\textbf{Approach} & $\bolds{\beta}_{0}$ & $\bolds{\beta}_{1}$ & $\bolds{\beta}_{2}$ & $\bolds{\beta}_{3}$ & $\bolds{\beta}_{4}$\\
\hline
Bayes & 0.635 & 1.583 & 3.584 & 4.382 & 1.579\\
Maximum likelihood & 1.811 & 1.817 & 3.299 & 4.402 & 1.311\\
\hline
\end{tabular}
\end{table}

\begin{figure}

\includegraphics{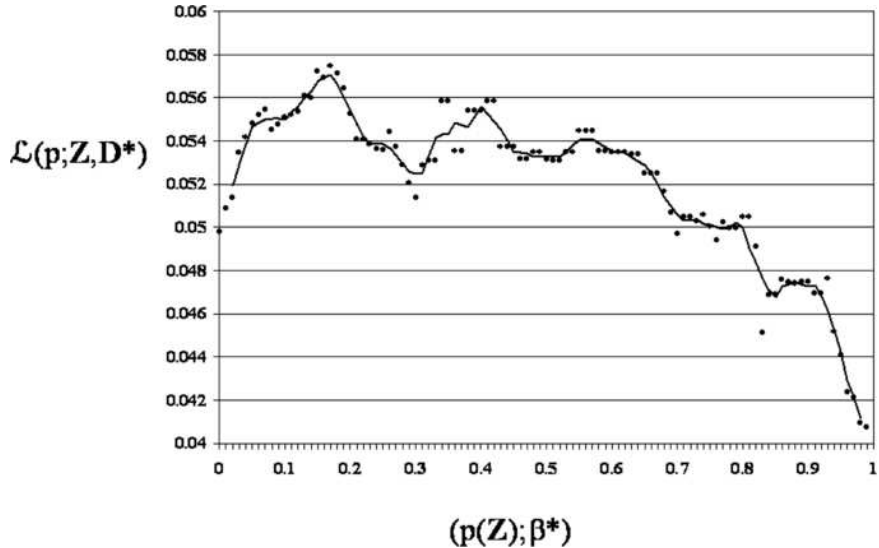}

\caption{Monte Carlo induced likelihood function.}
\label{fig5}\end{figure}

About 60 samples from the joint posterior distribution of $(\beta
_{0},\ldots ,\beta _{4})$ were generated, and for each sample, the quantity $%
[1+\exp (-\sum_{u=0}^{4}\beta _{u})]^{-1}$ computed. Here $\mathbf{Z}%
=(1,1,1,1)$, suggesting that the next crossing is to be on the new bridge
which is one mile away from all the four city centers of interest.
Associated with each generated sample is also the probability of the sample;
this is provided by the joint probability density. Figure~\ref{fig5} shows a plot of
the computed quantity mentioned above [our $(p(\mathbf{Z}),\bolds{\beta }%
^{\ast })$ of Section~\ref{sec4.3}] versus the joint probability. A smoothed plot,
smoothed by a moving average of five consecutive points, is the Monte Carlo
induced likelihood.

Since the new bridge has experienced 4 previous crossings and none of these
crossings have experienced an IED attack, $\mathbf{x}=(0,0,0,0)$; thus, $%
\mathcal{L}(p;\mathbf{x})=(1-p)^{\sqrt{2}}$, see equation (\ref{eq4.4}). With the
above in place, all the ingredients needed to compute $P(X=1;\mathbf{x;Z,D}%
^{\ast })$---equation (\ref{eq4.7})---are at hand, save for $\pi (p)$ the prior.
Supposing $\pi (p)$ uniform on $(0,1)$, we have%
\[
P(X=1;\mathbf{x;Z,D}^{\ast })\varpropto \int_{0}^{1}p(1-p)^{\sqrt{2}}%
\mathcal{L}(p;\mathbf{Z,D}^{\ast })\,dp,
\]%
with $\mathcal{L}(p;\mathbf{Z,D}^{\ast })$ given by the likes of Figure~\ref{fig5}.
This can be numerically evaluated for a range of $p$, say, $p=0.05,0.1,\ldots
,0.95,1$, to obtain $P(X=1;\break\mathbf{x,Z,D}^{\ast })\varpropto 0.129$.
Similarly, we obtain $P(X=0;\mathbf{x,Z,D}^{\ast })\varpropto 0.293$. The
normalizing constant is $0.422$, giving $P(X=1;\mathbf{x,Z,D}^{\ast })=0.306$
and $P(X=0;\mathbf{x,Z,D}^{\ast })=0.694$. Thus, the probability of
encountering an IED on the next crossing on the ``new
bridge'' is 0.306.

\subsubsection{Optimal route selection for logistical network}\label{sec5.2.1}

In order to prescribe an optimal route for the network of Figure~\ref{fig1}, we need
to calculate the probability of encountering an IED on each of the remaining
9 links of the network in a manner akin to that given above for link 9, the
``new bridge.''   This requires that we have
the vectors $\mathbf{x}$ and $\mathbf{Z}$ for each of these links, where~$%
\mathbf{x}$ is the historic IED experience for a link, and $\mathbf{Z}$ is
the vector of covariates associated with the links. This we do not have and
are unable to obtain for reasons of security. Consequently, and purely with
the intent of illustrating how our decision theoretic framework can be put
to work, we shall make some meaningful specifications about the $p(j)$'s, $%
j=1,\ldots ,8,10$. These will be based on the relative lengths of each link,
relative to the length of link 9 for which $p(9)$ has been assessed as
0.306; that is, calibrate the required $p(j)$'s in terms of $p(9)$.

To do the above, we start by remarking that links 1 and 2 are of almost
equal length, and are about two-thirds the length of link 9. Links 3 to 8
are of equal length and are about one-fifth the length of link 9, whereas
link 10 is about half the length of link 9. Note that Figure~\ref{fig1} is not drawn
to scale. Thus, we set $p(1)=p(2)=(0.66)(0.306)=0.20$, $%
p(10)=(0.50)(0.306)=0.15$ and $%
p(3)=p(4)=p(5)=p(6)=p(7)=p(8)=(0.20)(0.306)=0.06$. These choices are purely
illustrative; we could have used other methods of scaling such as the
logarithmic or the square root.

In addition to specifying the $p(j)$'s, we also need to specify utilities.
For this we propose a utility function of the form $1-n/x$ for a successful
route traversal. Here $n$ is the number of links in the route, and $x$ is a
constant which ensures that a successful traversal does not result in a
negative utility. Specifically, the idea here is that a successful traversal
yields a utility of one, but each link in the route contributes to a
disutility to which is assigned a~weight $1/x$. Choice $D_{1}$ entails the
route $(1,2,9)$ and with $x$ chosen to be 100, the utility of a successful
traversal on this route will be $1-3/100$. Similarly, the failure to achieve
a successful traversal yields a utility of $0-n/x$, yielding a negative
utility of $-n/x$, which in the case of route $(1,2,9)$ with $x=100$ is $%
-3/100$.

The above choices for utility do not take into consideration things such as
composition of the convoys, traversal time, vicinity to hostile territory,
costs of disruption, etc. With the above in place, and assuming independence
of the IED placement events, it can be easily seen that the expected
utilities of choices $D_{1}$, $D_{2}$, and $D_{3}$ are 0.414, 0.361, and
0.430, respectively. Thus, for the given choices of probabilities and
utilities, $\mathcal{D}$'s optimal route will be $D_{3}$, which is $%
(1,2,3,4,10)$. Observe that neither the shortest nor the longest routes are
optimal. Sensitivity of $\mathcal{D}$'s final choice to values of $x$ other
than 100 can be explored. For example, were $x$ taken to be 10, then $D_{1}$
will turn out to be $\mathcal{D}$'s optimal choice. This is because it turns
out the probability of a successful traversal via choices $D_{1}$, $D_{2}$,
and $D_{3}$ turns out to be rather close to each other, namely, 0.444, 0.441,
and 0.480, respectively.

This completes our discussion on illustrating the workings of the proposed
approach vis-\`{a}-vis the network of Figure~\ref{fig1}, and closes the
paper.

\section*{Acknowledgments}
The author was exposed to the IED problem by Professors Robert Koyak, Lynn
Whittaker, and (Col.) Alejandro Hernandez of the Naval Postgraduate School
(NPS), in Monterey, CA. Joshua Landon's help with the computations and
simulations of Section~\ref{sec5.1} is deeply acknowledged. Anna Gordon painstakingly
generated the data of Table~\ref{tab1}, whose source was made available to us by Dr.
Robert Bonneau of the Air Force Office of Scientific Research. The several
helpful comments by the referees, the Editor, Professor Fienberg, and the
Fienberg-Thomas paper have enabled the author to cast the problem of route
selection in a broader context. Work on this paper began when the author was
a visitor at NPS during the summer of 2008.


\printaddresses

\end{document}